\documentstyle[12pt,epsf,rotating]{article}  
\def\fo{\hbox{{1}\kern-.25em\hbox{l}}}

\def\bibitemx#1{\bibitem{#1}}
\def\hbh{h^0_{\rm bh}}
\def\hew{h^0_{\rm ew}}
\def\hu{h^0_u}
\def\hsm{h^0_{\rm sm}}
\def\hlight{h^0_{\rm light}}
\def\hheavy{h^0_{\rm heavy}}

\def\beq{\begin{equation}}
\def\eeq{\end{equation}}
\def\eq{\end{equation}}
\def\to{\rightarrow}

\def\bsg{\ifmmode B\to X_s\gamma\else $B\to X_s\gamma$\fi}
\def\bsll{\ifmmode B\to X_s\ell^+\ell^-\else $B\to X_s\ell^+\ell^-$\fi}
\def\bstt{\ifmmode B\to X_s\tau^+\tau^-\else $B\to X_s\tau^+\tau^-$\fi}
\def\shat{\ifmmode \hat{s}\else $\hat{s}$\fi}

\newcommand{\newc}{\newcommand}

\newc{\lcal}{\int {\cal L}dt}
 
\newc{\LSP}{{\chi^0_1}}
\newc{\stauR}{{\tilde \tau_R}}
\newc{\stau}{{\tilde \tau_1}}
\newc{\mstop}{m_{\tilde{t}}}
\newc{\mHpm}{m_{H^\pm}}
\newc{\gsim}{\lower.7ex\hbox{$\;\stackrel{\textstyle>}{\sim}\;$}}
\newc{\lsim}{\lower.7ex\hbox{$\;\stackrel{\textstyle<}{\sim}\;$}}
\newc{\ie}{{\it i.e.}}          
\newc{\etal}{{\it et al.}}
\newc{\eg}{{\it e.g.}}          
\newc{\kev}{\hbox{\rm\,keV}}            
\newc{\mev}{\hbox{\rm\,MeV}}            
\newc{\gev}{\hbox{\rm\,GeV}}            
\newc{\tev}{\hbox{\rm\,TeV}}
\newc{\xpb}{\hbox{\rm\, pb}}
\newc{\xfb}{\hbox{\rm\, fb}}

\newc{\mtop}{m_t}
\newc{\mbot}{m_b}
\newc{\mz}{m_Z}
\newc{\mw}{M_W}
\newc{\alphasmz}{\alpha_s(m_Z^2)}
\newc{\swsq}{\sin^2\theta_W}
\newc{\tw}{\tan\theta_W}
\newc{\cw}{\cos\theta_W}
\newc{\sw}{\sin\theta_W}
\newc{\BR}{\hbox{\rm BR}}
\newc{\zbb}{Z\to b\bar}
\newc{\Gb}{\Gamma (Z\to b\bar b)}
\newc{\Gh}{\Gamma (Z\to \hbox{\rm hadrons})}
\newc{\rbsm}{R_b^\hbox{\rm sm}}
\newc{\rbsusy}{R_b^\hbox{\rm susy}}
\newc{\drb}{\delta R_b}

\newc{\sgn}{\mbox{sgn}}
\newc{\tbeta}{\tan\beta}
\newc{\uL}{{\tilde u_L}}
\newc{\uR}{{\tilde u_R}}
\newc{\cL}{{\tilde c_L}}
\newc{\cR}{{\tilde c_R}}
\newc{\tL}{{\tilde t_L}}
\newc{\tR}{{\tilde t_R}}
\newc{\dL}{{\tilde d_L}}
\newc{\dR}{{\tilde d_R}}
\newc{\sL}{{\tilde s_L}}
\newc{\sR}{{\tilde s_R}}
\newc{\bL}{{\tilde b_L}}
\newc{\bR}{{\tilde b_R}}
\newc{\eL}{{\tilde e_L}}
\newc{\eR}{{\tilde e_R}}
\newc{\mhp}{m_{H^\pm}}
\newc{\mhalf}{m_{1/2}}
\newc{\emt}{{e/\mu /\tau}}

\newc{\lR}{\tilde{l}_R}
\newc{\lL}{\tilde{l}_L}
\newc{\nL}{\tilde{\nu}_L}
\newc{\na}{\chi^0_1}
\newc{\nb}{\chi^0_2}
\newc{\nc}{\chi^0_3}
\newc{\nd}{\chi^0_4}
\newc{\ca}{\chi^{\pm}_1}
\newc{\cb}{\chi^{\pm}_2}
\newc{\camp}{\chi^\mp_1}
\newc{\cbmp}{\chi^\mp_1}
\newc{\capos}{\chi^{+}_1}
\newc{\caneg}{\chi^{-}_1}
\newc{\phit}{\phi_t}
\newc{\phib}{\phi_b}
\newc{\phiew}{\phi_{ew}}
\newc{\htz}{h^0_t}
\newc{\hbz}{h^0_b}
\newc{\hewz}{h^0_{ew}}
\newc{\hsmz}{h^0_{sm}}
\newc{\huz}{h^0_u}
\newc{\hsusyz}{h^0_{susy}}

\def\mp{M_P}

\def\beq{\begin{equation}}
\def\eeq{\end{equation}}
\def\bea{\begin{eqnarray}}
\def\eea{\end{eqnarray}}
\def\slashchar#1{\setbox0=\hbox{$#1$}           % set a box for #1
   \dimen0=\wd0                                 % and get its size 
   \setbox1=\hbox{/} \dimen1=\wd1               % get size of /
   \ifdim\dimen0>\dimen1                        % #1 is bigger 
      \rlap{\hbox to \dimen0{\hfil/\hfil}}      % so center / in box 
      #1                                        % and print #1
   \else                                        % / is bigger 
      \rlap{\hbox to \dimen1{\hfil$#1$\hfil}}   % so center #1
      /                                         % and print /
   \fi}                                         %
\catcode`@=11
\long\def\@caption#1[#2]#3{\par\addcontentsline{\csname
  ext@#1\endcsname}{#1}{\protect\numberline{\csname
  the#1\endcsname}{\ignorespaces #2}}\begingroup
    \small
    \@parboxrestore
    \@makecaption{\csname fnum@#1\endcsname}{\ignorespaces #3}\par
  \endgroup}
\catcode`@=12

\def\jfig#1#2#3{
 \begin{figure}
 \centering
 \epsfysize=3.0in
 \epsffile{#2}
 \caption{#3}
 \label{#1}
 \end{figure}}
\begin{document}

\baselineskip=18pt

\begin{titlepage}
\begin{flushright}
UCD-2000-2 \\
hep-ph/0001226  
\end{flushright}

\begin{center}
\vspace{1cm}

{\Large \bf
Detecting a light Higgs boson at the Fermilab
Tevatron through enhanced decays to photon pairs}

\vspace{0.5cm}

{\bf Stephen~Mrenna and James Wells}

\vspace{.8cm}

{\it Davis Institute for High Energy Physics \\
   University of California, Davis, CA 95616}

\end{center}
\vspace{1cm}

\begin{abstract}
\medskip

We analyze the prospects of the Tevatron for finding a Higgs boson
in the two photon decay mode. 
We conclude that the Standard Model (SM) Higgs boson will likely not be 
discovered in this mode.  However,
we motivate several theories beyond the SM, including the MSSM, that predict
a Higgs boson with enhanced branching fractions into photons, and calculate
the luminosity needed to discover a general Higgs boson at the Tevatron
by a two-photon invariant mass peak at large transverse momentum.
We find that a high luminosity Tevatron will play a significant role
in discovering or constraining these theories.

\end{abstract}

\bigskip
\bigskip

\end{titlepage}

%%%%%%%%%%%%%%%%%%%%%%%%%%%%%%%%%%%%%%%%%%%%%%%%%%%%%%%%%%%%%%%

\section*{Introduction}

In this letter, we investigate the possibility that
the Higgs boson(s) $h$ associated with electroweak symmetry breaking may
be found in the $h\to \gamma\gamma$ decay channel at the Fermilab
Tevatron.
Our intention is to augment the many important studies 
preceding and associated with the RunII workshop on Supersymmetry
and Higgs physics at the Tevatron~\cite{runII web}.  
In these studies, Standard Model
Higgs boson detectability
has been studied vigorously in its most promising production and decay
channels.  The classic channel of $p\bar p\to Wh\to l\nu b\bar b$ remains
the most important channel in the search for the SM Higgs boson, yet
other modes can contribute to the total signal significance and perhaps yield
evidence for the Higgs boson if sufficient luminosity is attained.

We wish to study in detail the Higgs boson decays to two photons
for many reasons.  First, in our estimation
this decay mode has not received adequate
attention in previous studies.  The capabilities of
Higgs boson discovery in this mode should be carefully documented
in order to better understand the Tevatron's full potential for
Higgs boson detection.   Second, there are many interesting and motivated
theories that predict an enhanced decay rate into
the $\gamma\gamma$ channel,
and simultaneous suppression of the $h\to b\bar b$ channel.  Therefore,
in these cases, non-standard search strategies must be employed to either
find this Higgs boson or rule out its existence in  the
kinematically accessible mass range.  And finally, we feel that studies
such as these contribute to a more knowledgeable discussion
regarding the worth of a higher luminosity Tevatron ({\it e.g.}, run III).

Since there is no renormalizable and gauge invariant
operator in the Standard Model that leads to $h\to \gamma\gamma$ decays,
it must be induced by electroweak symmetry breaking effects.
The decay proceeds mainly through loop diagrams containing
$W^\pm$ bosons and the $t$ quark.  The $W^\pm$ boson loop
is dominant.  The branching fraction for this decay in the light Higgs boson
mass range $100\lsim m_h\lsim 150\gev$  is never much larger than 
$10^{-3}$ since
the $\gamma\gamma$ partial width must compete with the larger partial
widths associated with $b\bar b$, $\tau^+\tau^-$, $c\bar c$, gg, and 
$WW^*$ decays.
The branching fractions for the Standard Model Higgs boson have been
reliably calculated in Ref.~\cite{Djouadi:1998yw}.  The maximum of the Standard
Model branching fraction is 0.22\% and is reached at $m_h= 125\gev$.  For
$m_h>125\gev$, the branching fraction
falls somewhat rapidly due to the increased importance of $WW^*$ decays.
\bigskip

%%%%%%%%%%%%%%%%%%%%%%%%%%%%%%%%%%%%%%%%%%%%%%%%%%%%%%%%%%%%%%%%%%%
\section*{Beyond the Standard Model}
\bigskip
 
It is widely recognized that the Standard Model is an unsatisfactory
explanation of electroweak symmetry breaking.    In this
section, we review several well-motivated alternatives to the Standard
Model Higgs sector.  The first of these is low--scale supersymmetry,
where the symmetry breaking tasks are shared by two fields: $H_u$
and $H_d$. $H_u$ receives a vacuum expectation value and gives mass to the
up-type quarks, while $H_d$ receives a vacuum expectation value and
gives mass to the down-type quarks.  Both $\langle H_u\rangle$ and 
$\langle H_d\rangle$ vacuum expectation values
contribute to the $W^\pm$ and $Z^0$ masses.
In general, the sharing of the electroweak symmetry breaking task 
between two or
more fields will disrupt expectations of Higgs boson phenomenology
based solely on the analysis of the Standard Model Higgs boson.
It is important to identify regions of parameter space where our
naive expectations fail, and where a more expansive search strategy
must be engaged to find evidence of a Higgs boson.  

The mass matrix for the CP-even neutral Higgs bosons of supersymmetry 
in the $\{ H^0_d, H^0_u \}$ interaction basis is
\beq
{\cal M}^2 = \left( \begin{array}{cc}
   m^2_A \sin^2\beta +m^2_Z \cos^2\beta & -\sin\beta\cos\beta (m^2_A+m^2_Z) \\
   -\sin\beta\cos\beta (m^2_A+m^2_Z) & m^2_A\cos^2\beta +m^2_Z\sin^2\beta 
   \end{array}\right) + \left( \begin{array}{cc}
     \Delta_{dd} & \Delta_{ud} \\
     \Delta_{ud} & \Delta_{uu} \end{array} \right),
\eeq
where $m^2_A$ represents the pseudo-scalar mass, whose value is set by 
supersymmetry breaking, and $\Delta_{ij}$ are quantum corrections whose
form can be extracted from Ref.~\cite{deltas}.

In the limit $m_A\gg m_Z$ the mass eigenstates of the above mass matrix
are
\bea
\hlight & = & \cos\beta H^0_d+\sin\beta H^0_u \\
\hheavy & = & -\sin\beta H^0_d + \cos\beta H^0_u.
\eea
One can immediately see that 
$\langle\hlight\rangle =v$ and $\langle\hheavy\rangle=0$, and
it is also true that
all interactions of $\hlight$ are equivalent to the SM Higgs boson.
It is instructive to rotate the Higgs mass matrix to the 
$\{ \hlight,\hheavy \}$ basis:
\beq
{\cal M'}^{2}= \left( \begin{array}{cc}
   m^2_Z\cos^2 2\beta & -m^2_Z\sin 2\beta \cos 2\beta \\
   -m^2_Z\sin 2\beta \cos 2\beta & m^2_A +m^2_Z\sin^2 2\beta 
   \end{array} \right) +\left( \begin{array}{cc}
     \Delta'_{11} & \Delta'_{12} \\
     \Delta'_{12} & \Delta'_{11} \end{array} \right),
\eeq
where the $\Delta'_{ij}$ can be expressed in terms of the more commonly
given corrections $\Delta_{ij}$,
\bea
\Delta'_{11} & = & \Delta_{dd}\cos^2\beta +2\Delta_{ud}\cos\beta\sin\beta
  +\Delta_{uu}\sin^2\beta \nonumber \\
\Delta'_{12} & = & -\Delta_{dd}\cos\beta\sin\beta +\Delta_{ud}\cos 2\beta
   +\Delta_{uu}\cos\beta\sin\beta \nonumber \\
\Delta'_{22} & = & \Delta_{dd}\sin^2\beta -2\Delta_{ud}\cos\beta\sin\beta
   +\Delta_{uu}\cos^2\beta . \nonumber
\eea
``Higgs decoupling'' in supersymmetry means that 
one Higgs boson stays light and couples just
like the SM Higgs boson as supersymmetry breaking
mass scales get very high.  This property of the supersymmetric Higgs sector
can be immediately understood as a complete $SU(2)$ Higgs doublet
becoming very heavy ($\hheavy, A^0, H^\pm$), while another doublet stays
light ($\hlight, Z^0_L, W^\pm_L$).  In the expressions above, this is
equivalent to noting that $m^2_A$  
occurs only in the ${\cal M'}^2_{22}$-element of the
$\hlight -\hheavy$ mass matrix.

In supersymmetry model building, the supersymmetry breaking scale is 
a free parameter and is cycled over a very large range.  
This gives the false impression that over the vast majority of
the parameter space, $m_A$ is sufficiently larger than $m_Z$ to be
in the ``decoupling region'' described in the previous two paragraphs,
and the lightest Higgs boson is well approximated by  $\hlight$.
However, a natural electroweak potential --- meaning a potential that has
no large cancellations to produce the $Z$ boson mass --- prefers
supersymmetry breaking near the weak scale.  If we take naturalness
and fine tuning arguments seriously, we expect $m_A\sim m_Z$, which leads to
potentially 
significant deviations of the light Higgs boson couplings to the SM particles.

An interesting departure from SM Higgs phenomenology
occurs when the light Higgs boson mass
eigenstate of supersymmetry is the weak eigenstate 
$h^0_u$~\cite{Baer:1998cm}-\cite{Carena:1999bh}.
This scenario, or close approximations to it, can naturally occur
in theories with large $\tan\beta =\langle H_u\rangle/\langle H_d\rangle$,
which are motivated by supersymmetric $SO(10)$ 
unification~\cite{Hall:1994gn}, and by minimal gauge-mediated 
supersymmetry theories that solve
the soft CP-violating phase problem~\cite{Dine:1997xk}.  
The $h^0_u$
eigenstate has no tree-level
coupling to $b\bar b$ or $\tau^+\tau^-$, and the total
width for this light Higgs boson is greatly reduced.  Loop corrections
can modify these arguments.  For example,
supersymmetry breaking can induce couplings
such as $\lambda_b' H_u^*b\bar b$ (and $\lambda_\tau' H_u^*b\bar b$) 
in addition to the usual
$\lambda_b H_db\bar b$.  The most important of these corrections often
comes from gluino-squark loops (which do not contribute to $\lambda_\tau'$).

If a significant $\lambda_b'$ coupling is induced, the condition
for shutting off
the $b\bar b$ coupling is to shift the Higgs rotation angle
to $\tan\alpha =\epsilon/\tan\beta$ (or $\tan\alpha=-\epsilon/\tan\beta$ 
for the
case when the heavier CP even Higgs boson is Standard Model--like) where 
$\epsilon \equiv\Delta m_b/(m_b-\Delta m_b)$ and
$\Delta m_b\equiv \lambda_b' \langle H_u\rangle$.  The $\tau\tau$ branching is
not zero now, but is modified by a factor of $\epsilon^2$ compared
to the Standard Model (in the limit that $\lambda_\tau'$ is small). 
In contrast to the suppressed down-type fermion
couplings,
the partial width to two photons is equal to that of the Standard
Model since no down-type quarks or leptons contribute significantly
to the loop diagrams in either case.  
For these reasons, the branching ratio for the two photon final state
can be greatly enhanced.  Furthermore, the
production rates through $gg\to \hu$ and $q\bar q'\to W\hu$ are the
same as in the Standard Model, since neither of these rely on
the down-type fermion couplings.

Other interesting theories imply enhanced branching fractions
to two photons.  The bosonic Higgs $\hbh$ that gives all~\cite{bosonic higgs}, 
or rather
nearly all, the mass to the vector bosons and has no couplings to
fermions is a good example of a Higgs boson with enhanced branching
fractions to two photons.  However, the production cross-section of
$gg\to \hbh$ is negligible in this model since the top quark does not
couple to this Higgs.  One must rely completely on electroweak boson
couplings for production of the $\hbh$, such as in $q\bar q'\to W\hbh$ or
$WW\to \hbh$.

Another example that has suppressed couplings to the fermions is
an electroweak Higgs boson $\hew$ added to top-quark condensate 
models~\cite{Wells:1997pg}.
In this approach, the top and bottom quarks are assumed to get their
masses through a strongly coupled group that condenses top quark 
pairs~\cite{top condensation},
and all the remaining fermions and vector bosons get mass mainly through
$\langle \hew\rangle$.  A good approximation in studying the
phenomenology of a light $\hew$ is to assume that it couples like
the Standard Model Higgs to all particles except
the top quark and bottom quark, to which it has zero couplings.

In Fig.\ref{bryy} we plot the branching fraction into two photons for the
four Higgs bosons that we mentioned above: $\hsm$, $\hu$, $\hbh$,
and $\hew$.
%%%%%
\jfig{bryy}{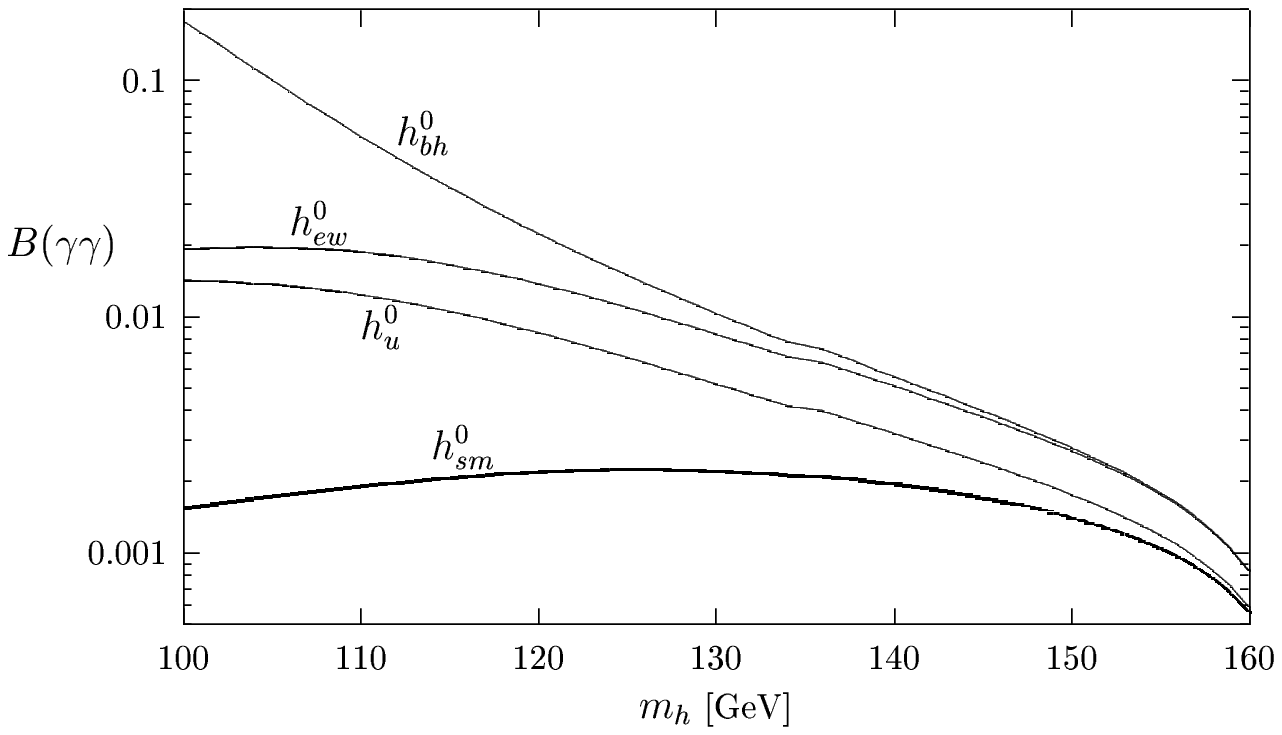}{Branching fraction into two photons for
four different types of Higgs bosons: (1) $\hsm$ is the Standard Model
Higgs boson, (2)  $\hu$ is the Higgs boson
with Yukawa couplings only with up--type fermions, which can be
a mass eigenstate in large
$\tan\beta$ supersymmetric theories, (3) $\hew$ is the Higgs boson 
that may help complete top quark condensation models as described
in the text, and (4)
$\hbh$ is a Higgs boson with tree 
level couplings only to $W$ and $Z$ bosons.}  
In each non-SM case considered, the branching fraction is larger than 
that of $\hsm$.  Some models that we have not discussed here may
have even higher branching fraction or perhaps lower.  It should be
kept in mind that {\em any} model of physics beyond the simple
Standard Model will likely have different branching fractions into two
photons.  Since the two photon partial width is a one-loop process,
it will also be sensitive
to new particles in loop diagrams.  Hence, even greater variability
is possible than what we have shown here.   For example, supersymmetric
partners in the loops may increase or decrease the overall partial width
of $h\to \gamma\gamma$~\cite{Spira:1995rr,Kane:1996ek,Djouadi:1998az}.  
In general, we should be prepared to discover
and study a Higgs boson with any branching fraction to two photons,
since that is perhaps the most likely branching fraction to be
altered significantly by new physics.

There are several sizable sources of Higgs boson production within
the Standard Model.  At the Tevatron, 
they are $gg\to h$, which is the largest, followed
by $q\bar q'\to Wh$ and $q\bar q\to Zh$.  For a heavy enough Higgs boson, 
the vector boson
fusion processes $WW,ZZ\to h$ are also competitive.
Although the decay branching fraction
of $h\to\gamma\gamma$ is of order $g^2\over 16\pi^2$, 
there is hope that the narrow $M_{\gamma\gamma}$
peak of the signal can be utilized to cut extraneous two photon backgrounds
to sufficiently low levels that a signal can be detected.
In the following two sections, we discuss search strategies based on
inclusive and exclusive final states.  A description
of our calculational methods
is provided in the Appendix.

%%%%%%%%%
\section*{Inclusive $\gamma\gamma +X$ production}

First, we consider the total inclusive production of Higgs bosons,
followed by their prompt decay to $\gamma\gamma$, where all
Higgs boson production mechanisms can contribute.  
%The background is dominated by the process $q\bar q\to \gamma\gamma$, with additional
%contributions from $\gamma\gamma +{\rm jets}$.  The jets in the event
%can cause additional worries. For example, the jet energies can be mismeasured
%resulting in spurious missing energy, or the jets can fake electrons.
%These fake backgrounds will play an important role in our comparison of
%signal to background.
To study inclusive
production, we apply cuts only on the properties of the individual
photons or the photon pair, without studying the rest of the event in
great detail.  Before applying any cuts, the photon energy $E^\gamma$ 
is smeared
by a resolution function typical of Run I 
conditions~\cite{experiment}:
\beq
{ \Delta E^\gamma \over E^\gamma }= \frac{.15}{\sqrt{E^\gamma\, (\gev )}}\oplus .03.
\label{photonres}
\eeq

To optimize the acceptance of signal events,
while reducing the ``irreducible'' backgrounds and 
those from jets fragmenting to photons, we
apply the cuts,
\bea
p_T^\gamma > 20~{\rm GeV}, |\eta^\gamma| <2 &{\rm (triggering~and~acceptance)}  \nonumber \\
\Delta R^{\gamma\gamma}\equiv\sqrt{(\eta_1^\gamma-\eta_2^\gamma)^2
+(\phi_1^\gamma-\phi_2^\gamma)^2}>0.7 & {\rm (separation)} \nonumber \\
\sum_{(i),R<.4}^{ } E_T^{(i)} - p_T^\gamma < 2~{\rm GeV} & {\rm (isolation)} .
%\nonumber
\label{cuts1}
\eea
The high $p_T$, central photons constitute a suitable trigger.
We failed to find a more efficient $p_T^\gamma$ cut
than the one listed.
With these cuts, the dominant source of background comes from the
$q\bar q\to \gamma\gamma$ process.  For each event, we treat the
diphoton pair as a Higgs boson candidate with mass $M_{\gamma\gamma}$.

A further cut on the angular distribution of the photons in the rest
frame of the Higgs boson candidate increases $S/B$:
\bea
|\cos\theta^*|<0.7.
\label{cuts2}
\eea
The angle $\theta^*$ is defined to be the angle that the photon makes with
the {\it boost} direction in the $\gamma\gamma$ rest frame.  The signal
is rather flat in $\cos\theta^*$ whereas the raw background peaks at
$|\cos\theta^*| =1$.  This cut is somewhat redundant to the
other acceptance cuts, but will suppress fake backgrounds.

With the above cuts it would require well over $100\xfb^{-1}$ of integrated
luminosity to even rule out a SM Higgs boson (95\% C.L.) at any mass
(the details will be given later).  
Therefore, new physics that provides a significant enhancement of the 
$\gamma\gamma +X$ total rate is required
for this kind of signal process to be a relevant search.  
Large enhancements can occur either in the production cross-sections
or in the decay branching fraction to photons.  
In the Standard Model, the $gg\to h$ process constitutes roughly $2/3$rds of
the total production rate up to about $m_h=160$ GeV, while the rest of the rate
is mainly $q\bar q\to W/Z+h$ production.
One does not expect the production cross-sections $q\bar q\to W/Z+h$ to ever
greatly exceed the SM production cross-section given the nature of Higgs
boson couplings to electroweak vector bosons.  One does expect, however,
that the $gg\to h$ rate could be greatly enhanced by an increased coupling
of the top quarks to Higgs boson~\cite{Spira:1997ce}, or 
by many virtual states contributing to the one--loop, 
effective $ggh$ coupling, or from higher dimensional operators
induced in theories with large extra dimensions~\cite{Hall:1999fe}.  
In fact, these effects that increase the
rate of $gg\to h$ production will usually also alter
the $h\to\gamma\gamma$ branching fraction.  Therefore, we focus
on the possibility of large enhancements of the ratio
\beq
R_{gg} = \frac{\sigma(gg\to h)B(h\to\gamma\gamma)}
         {\sigma(gg\to h)_{\rm sm}B(h\to\gamma\gamma)_{\rm sm}}.
\eeq

In the following we will investigate the inclusive $\gamma\gamma+X$
rate from $gg\to h$ signal production alone.  Although this underestimates
the total cross-section by not taking into account the $W/Z+h$
and $WW,ZZ\to h$ 
contributions, it lends itself to easy generalizations of $R_{gg}\gg 1$
where there is hope to find a signal at reasonable luminosity and
where the other contributions are very small in comparison.  
Later on, we will see that
the $q\bar q \to W/Z+h$ signature alone lends itself to a useful,
complementary analysis based on exclusive final states.
In Table~\ref{gghyy} we list the total SM signal cross-section
($gg\to h$ only) and the differential background rate after all cuts have
been applied.  
\begin{table}
\begin{center}
\begin{tabular}{ccc} \hline\hline
$m_h$ [GeV] & $\sigma_{\rm sig}(\gamma\gamma +X)$ [fb]
         & $d\sigma_{\rm bkgd}/dM_{\gamma\gamma}$ [fb/GeV] \\ \hline
100 & 1.49 & 39.3 \\
110  & 1.43 & 27.6 \\
120 & 1.29 & 20.2 \\
130 & 1.02 & 15.1 \\
140 & 0.73 & 12.3 \\
150 & 0.42 & 9.8 \\
160 & 0.13 & 7.4 \\
170 & 0.029 & 5.7 \\
\hline \hline
\end{tabular}
\end{center}
\caption{The total $\gamma\gamma+X$ production rate (in fb) 
for a Standard Model Higgs boson,
and the differential rate (in fb per GeV) for the Standard Model
backgrounds after applying cuts Eqs.~(\ref{cuts1})-(\ref{cuts2}).
\label{gghyy}}
\end{table}

The Higgs boson width in the Standard Model is less than $20\mev$
for $m_h<150\gev$.  Therefore, 
the invariant mass measurement
of the two photons will have a spread entirely due to the photon energy
resolution of the detector, 
which we call $\Delta M_{\gamma\gamma}$.  In Table~\ref{myy}
we show $\Delta M_{\gamma\gamma}$ for various Higgs boson masses,
based on folding the photon energy resolution function Eq.~(\ref{photonres}) 
with the photon kinematics.
\begin{table}
\begin{center}
\begin{tabular}{c|cccccccc} \hline\hline
$m_h$ [GeV] & 100 & 110 & 120  & 130 & 140 & 150 & 160 & 170 \\
$\Delta M_{\gamma\gamma}$ [GeV] & 1.52 & 1.64 & 1.76 & 1.88 & 1.99 & 
         2.08 & 2.23 & 2.47 \\ \hline\hline
\end{tabular}
\end{center}
\caption{The invariant mass resolution for a narrow signal from 
our simulations.  
The resolution is the $1 \sigma$ value of a Gaussian fit to the 
simulated signal
after applying cuts Eqs.~(\ref{cuts1})-(\ref{cuts2}).
\label{myy}}
\end{table}

Based on Tables~\ref{gghyy} and~\ref{myy}, 
we are now able to determine the significance of the signal
with respect to background after all cuts.  We use the formula, 
\beq
N_S=\frac{S}{\sqrt{B}}=\frac{0.96\sigma_{\rm sig}
      \sqrt{{\cal L}}}{\sqrt{\hat \sigma_{\rm bkgd}}},
\label{nsigma}
\eeq
where
\beq
\hat \sigma_{\rm bkgd}=4\Delta M_{\gamma\gamma}\frac{d\sigma_{\rm bkgd}}
  {dM_{\gamma\gamma}},
\label{bkgd}
\eeq
and ${\cal L}$ is the integrated luminosity.
This formula counts the significance of signal to background within
a mass window
$M_{\gamma\gamma}\pm 2\Delta M_{\gamma\gamma}$.  
This is a conservative and simple choice.  When both the signal and background
can be described adequately using Gaussian statistics, and the signal itself
has a Gaussian shape, and the background is constant, the optimal mass window
is $M_{\gamma\gamma}\pm \sqrt{2}\Delta M_{\gamma\gamma}$.  In our case, the
background is not a constant, but the differential distribution is well 
approximated by a straight line with a negative slope.  Therefore, an asymmetric
mass window (with respect to the peak) 
would most likely yield the best significance.
We also require everywhere
in our analysis that no limit or discovery capability is possible unless
at least 5 events are present in this $2\sigma$ spread mass bin.  On the
graphs we show below, this is a limitation mainly for the $2\xfb^{-1}$
integrated luminosity curve.  
From Eqs.~(\ref{nsigma}) and (\ref{bkgd}), 
it is worth noting that an increase in integrated
luminosity is equivalent to an improved energy resolution.

In Fig.~\ref{gghyy2} we plot the $95\%$ C.L.\ ($N_S=1.96$)
exclusion curves for a given
luminosity in the $R_{gg}$-$m_h$ plane.  The SM
Higgs boson corresponds to $R_{gg}=1$ across the plot.
We have put on a line on the graph corresponding to 
$h_u^0$ to give a non-SM reference
example of $R_{gg}$.  Other theories such as those
discussed above can have
$R_{gg}$ much greater than that of $h_{\rm sm}^0$ 
or $h^0_{\rm u}$.  The plot is intended to be useful for comparing
any theory to Tevatron capabilities.
%%%%%%%%%%%%%%%%%%
\jfig{gghyy2}{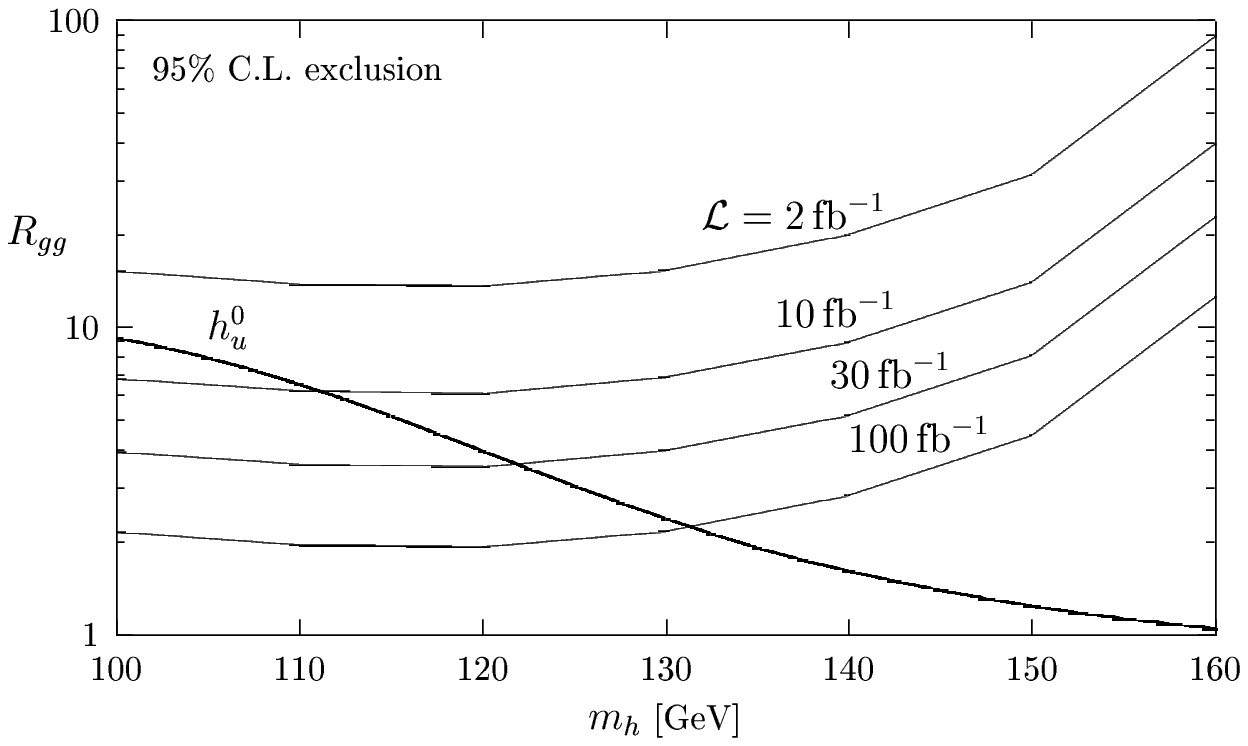}{95\% C.L. luminosity contours of Higgs boson
detection in the $R_{gg}$-$m_h$ plane.  For a given luminosity curve,
the region below the curve cannot be ruled out with Tevatron data.}

For a given integrated luminosity, the region above the corresponding
curve can be ruled out to $95\%$ confidence level.  Therefore, with
$30\xfb^{-1}$ one could exclude a $h^0_u$ up to $120\gev$.  The solid
lines never cross $R_{gg}=1$ which indicates that the SM Higgs boson could
not be excluded in the $\gamma\gamma$ mode
by the Tevatron even with over $100\xfb^{-1}$ of data.
One interesting limit to consider is a Higgs boson with only one--loop decays
to $gg$ and $\gamma\gamma$ final states.  In this case, the production cross
section $\times$ branching ratio is proportional to 
$\Gamma(h\to gg) BR(h\to\gamma\gamma)\simeq \Gamma(h\to\gamma\gamma)$,
and $R_{gg} = {\Gamma(h\to\gamma\gamma) \over \Gamma(h_{SM}\to gg)
BR(h_{SM}\to\gamma\gamma)} \simeq 10^3 {\Gamma(h\to\gamma\gamma) \over \Gamma(h_{SM}\to gg)}$.
Therefore, large values of $R_{gg}$ are not unreasonable.

Discovery of Higgs bosons with enhanced $\gamma\gamma+X$ 
production rates 
requires higher significance.  For $N_S=5$ we plot in Fig.~\ref{gghyy5} the
necessary enhancement $R_{gg}$ to see a signal at
this level at the Tevatron.  With less than $30\xfb^{-1}$, discovery
is not likely for a $h^0_u$ Higgs boson with mass greater than $100\gev$.
Therefore, Tevatron detection sensitivity in this channel is not as
good as the Higgs boson search capacity at LEP2, which should exceed
$105\gev$ for both $h^0_{\rm sm}$ and $h^0_{u}$.  Nevertheless 
other theories
with larger enhancements of $\sigma(gg\to h)B(h\to \gamma\gamma)$
may be discovered in the $\gamma\gamma +X$ mode first.
%%%%%%%%%%%%%%%%%%%%%%%%%%%%%%%%%%5
\jfig{gghyy5}{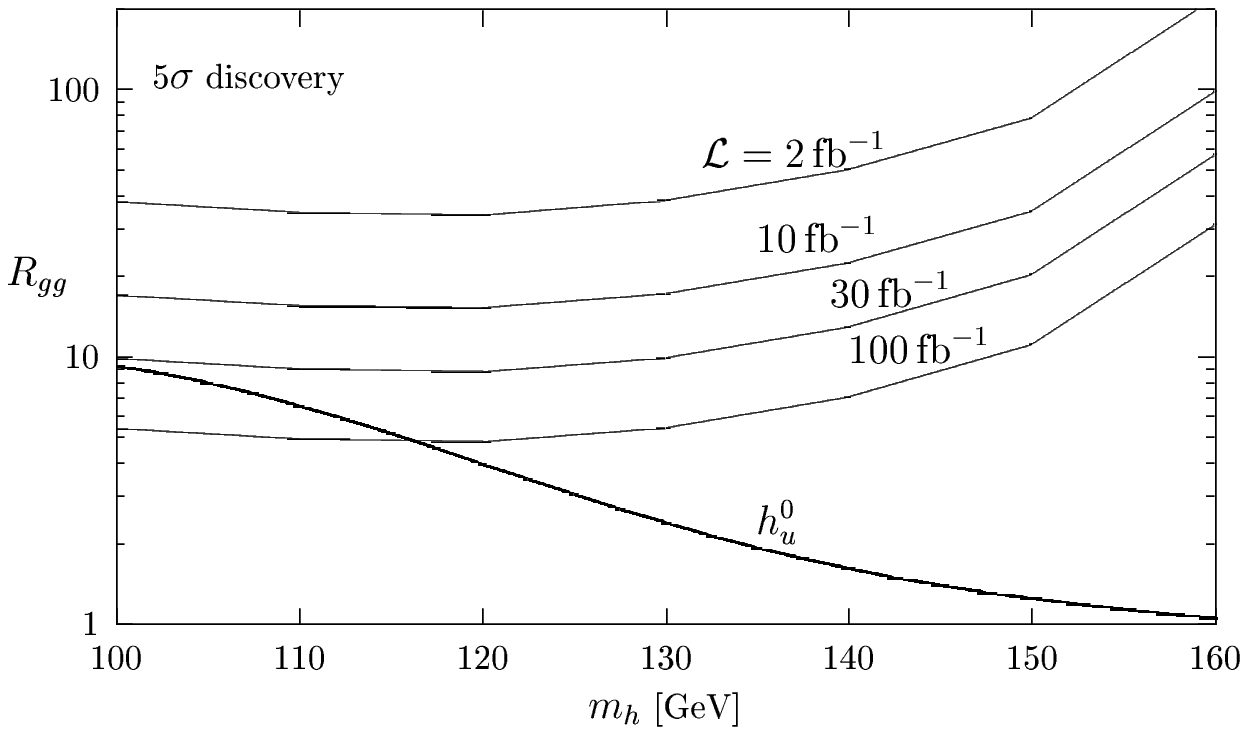}{$5\sigma$ discovery contours of Higgs boson
detection in the $R_{gg}$-$m_h$ plane for a given luminosity.  
For each luminosity curve,
the region above the curve can be discovered with Tevatron data.}

%%%%%%%%%%%%%%%%%%%%%%%%%%%%%%%%%%%%%%%%%%%%%%55
\section*{Exclusive $W/Z+h\to W/Z+\gamma\gamma$ signal of Higgs bosons}

We now attempt to gain more significance of signal to background
by employing additional cuts.  It is well known that the kinematics
of resonance production at hadron colliders can be significantly
affected by multiple soft gluon emission.
Because of the different color factors associated with the $q\bar
q\to\gamma\gamma$ and $gg\to h$ processes, the $p_T^{\gamma\gamma}$
spectrum of the Higgs boson signal is harder than the background.  One
strategy of LHC searches is to exploit this difference to establish a
Higgs signal~\cite{Abdullin:1998er}.
However, the process $W/Z+h\to W/Z+\gamma\gamma$ typically has
large
$p_T^{\gamma\gamma}$ even before including these QCD effects.
At the Tevatron collider, the $W/Z+h$
production process is relatively much more important than at the LHC,
and quickly becomes the dominant process at even moderate values of
$p_T^{\gamma\gamma}$ with respect to $M_{\gamma\gamma}$.
For this reason, and because many of the extensions of the SM
considered here have no $gg\to h$ component or only one of SM strength,
we concentrate only on the $W/Z+h$ signal in the following.  For
reasons discussed later, the $WW/ZZ \to h$ signal is not as relevant
for our analysis.
  
We have done an analysis of varying the $p_T^{\gamma\gamma}$ cut to
maximize the total signal significance.   
We  find that we optimally retain a significant portion of 
the total Higgs boson signal while reducing the
backgrounds with the requirement,
\beq
 p_T^{\gamma\gamma} > M_{\gamma\gamma}/2.
\eeq
Also, the two photons from the Higgs boson decay tend to be balanced
in $p_T$, so we demand
\beq
p_T^\gamma>M_{\gamma\gamma}/3
\eeq
as a further aid to reduce backgrounds and increase $S/B$.

After demanding such a significant cut on $p_T^{\gamma\gamma}$, the dominant
background becomes $\gamma\gamma$+1 jet.  However, the signal will most likely
{\it not} have this topology.  Rather, the decay of $W$ and $Z$ bosons can
lead to: (1) two hard jets with $M_{jj}\simeq M_W,M_Z$, (2) one or more
high $p_T$ leptons from $W\to e,\mu$ and $Z\to ee,\mu\mu$, or (3) missing
transverse energy from $Z\to\nu\nu$, $W\to\tau\to~{\rm soft~jet}$, and
$W\to e,\mu \to~{\rm soft~or~very~forward~leptons}$.
%Therefore, we employ a series of cuts to isolate the exclusive 
%process $W/Z(\to f\bar f)+h(\to\gamma\gamma)$.  
%In $Wh$ and $Zh$ production, the vector
%bosons can decay either
%leptonically or to jets, or to invisible neutrinos for $Z$ bosons.
Therefore, it is useful to consider 
$\gamma\gamma$ signals that have one or two leptons, {\it or}
missing energy, {\it or} two leading jets with $M_{j_1j_2}\simeq m_W,m_Z$.
To this end we require
at least one of the following ``vector boson acceptance'' 
          criteria to be satisfied:
\begin{itemize}
\item[(a)] $p_T^{e,\mu}>10\gev$ and $|\eta^{e,\mu}|<2.0$.
\item[(b)] $\slashchar{E_T}>20\gev$.
\item[(c)] 2 or more jets with $50\gev <M_{j_1j_2}<100\gev$.
\end{itemize}
To perform this analysis,
we constructed jets ($E_{T}^j>15\gev$, $|\eta^j|<2.5$ and $R=0.5$) using
the toy calorimeter simulation in {\sc PYTHIA} with an energy resolution
of $80\%/\sqrt{E^j(\gev )}$.  $\slashchar{E_T}$ was calculated by summing all
calorimeter cells out to $\eta=4$.

The vector boson acceptance
cuts eliminate a fair portion of the $\gamma\gamma$ plus
jet background, as well as a potential contribution from vector
boson fusion.  The total rate of the vector fusion process (without cuts)
is comparable to $W/Z+h$ only for $M_h > 160\gev$, where large
values of $B(h\to \gamma\gamma)$ are not well motivated
(see, {\it e.g.}, Fig.~\ref{bryy}). Nonetheless, we examined the
effects of replacing cuts (a)-(c) by the requirement 
$M_{j_1j_2}>100\gev$ to accept the jets associated with vector boson fusion:
$q\bar q\to q'\bar q'h$. The results
were not as promising as those
based on cuts (a)-(c), and so we did not include a $M_{j_1j_2}>100\gev$
acceptance cut in our analysis.

In Table~\ref{hyyV} we show the signal and differential cross-section
rates after all cuts, including the $p_T^{\gamma\gamma}>M_{\gamma\gamma}/2$
and ``vector boson acceptance'' requirements.
\begin{table}
\begin{center}
\begin{tabular}{ccc} \hline\hline
$m_h$ [GeV] & $\sigma_{\rm sig}(\gamma\gamma +V)$ [fb]
         & $d\sigma_{\rm bkgd}/dM_{\gamma\gamma}$ [fb/GeV] \\ \hline
100 & 0.134 & 0.102 \\
110  & 0.119 & 0.070 \\
120 & 0.097 & 0.047 \\
130 & 0.077 & 0.034 \\
140 & 0.049 & 0.026 \\
150 & 0.026 & 0.020 \\
160 & 0.0082 & 0.014 \\
170 & 0.0016 & 0.011 \\
\hline \hline
\end{tabular}
\end{center}
\caption{The total signal (in fb) for $\gamma\gamma +V$, where
$V$ represents additional states passing the ``vector boson acceptance''
criteria enumerated in the text.  The last column is the calculated
background given the same cuts.
\label{hyyV}}
\end{table}
Table~\ref{hyyV} can then be used to determine detectability of
a Higgs boson given its mass and $R_V$:
\beq
R_V = \frac{\sigma(W/Z+ h)B(h\to\gamma\gamma)}
       {\sigma(W/Z + h)_{\rm sm}B(h\to\gamma\gamma)_{\rm sm}}.
\eeq
The parameter $R_V$ is useful if we make the reasonable assumption 
that increases in
$\sigma(W+h)$ and $\sigma(Z+h)$ scale equivalently.

In Fig.~\ref{Vhyy2} we plot the $95\%$ C.L.\ ($N_S=1.96$)
exclusion curves for a given
luminosity in the $R_V$-$m_h$ plane.
On the curve we have put lines for $h_u^0$ and the purely gauge coupled
Higgs boson $h_{\rm bh}^0$. The SM
Higgs boson corresponds to $R_V=1$ across the plot.
For a given integrated luminosity, the region above the corresponding
curve can be ruled out to $95\%$ confidence level.  
The luminosity  curves
never cross $R_{V}=1$ which indicates that the SM Higgs boson could
not be excluded in the $\gamma\gamma$ mode
by the Tevatron even with $100\xfb^{-1}$ of data. However, with
$30\xfb^{-1}$ one could exclude $\hbh$ up to $137\gev$ and
$h_u^0$ up to $129\gev$ in the $\gamma\gamma$ channel alone.  
%$$$$$$$$$$$$$$$$$$$$$$$$$$$$$$$$$$$$$$$$$$$$$$$$$$$$$$$$$$
\jfig{Vhyy2}{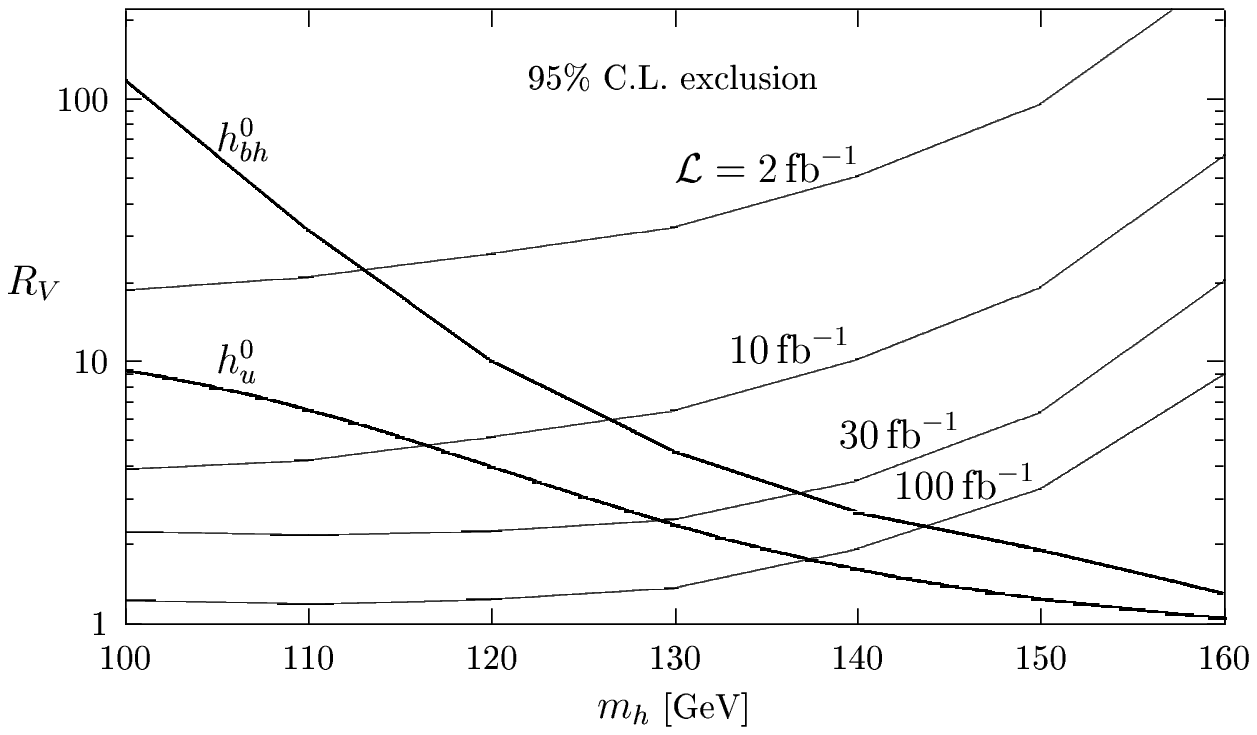}{95\% C.L. luminosity contours of Higgs boson
detection in the $R_{V}$-$m_h$ plane.  For a given luminosity curve,
the region below the curve cannot be ruled out with Tevatron data.}

For $N_S=5$ discovery we plot in Fig.~\ref{Vhyy5} the
necessary enhancement of $B(h\to \gamma\gamma)$ to see a signal at
this level at the Tevatron.  Discovery is possible up to $126\gev$ for
the bosonic Higgs boson as long as at least $30\xfb^{-1}$ is obtained,
and $h^0_u$ can be discovered up to approximately $114\gev$.
Both discovery reaches are beyond the expected reach capacity of LEPII.
%$$$$$$$$$$$$$$$$$$$$$$$$$$$$$$$$$$$$$$$$$$$$$$$$$$$$$$$$$$$$$$$$$
\jfig{Vhyy5}{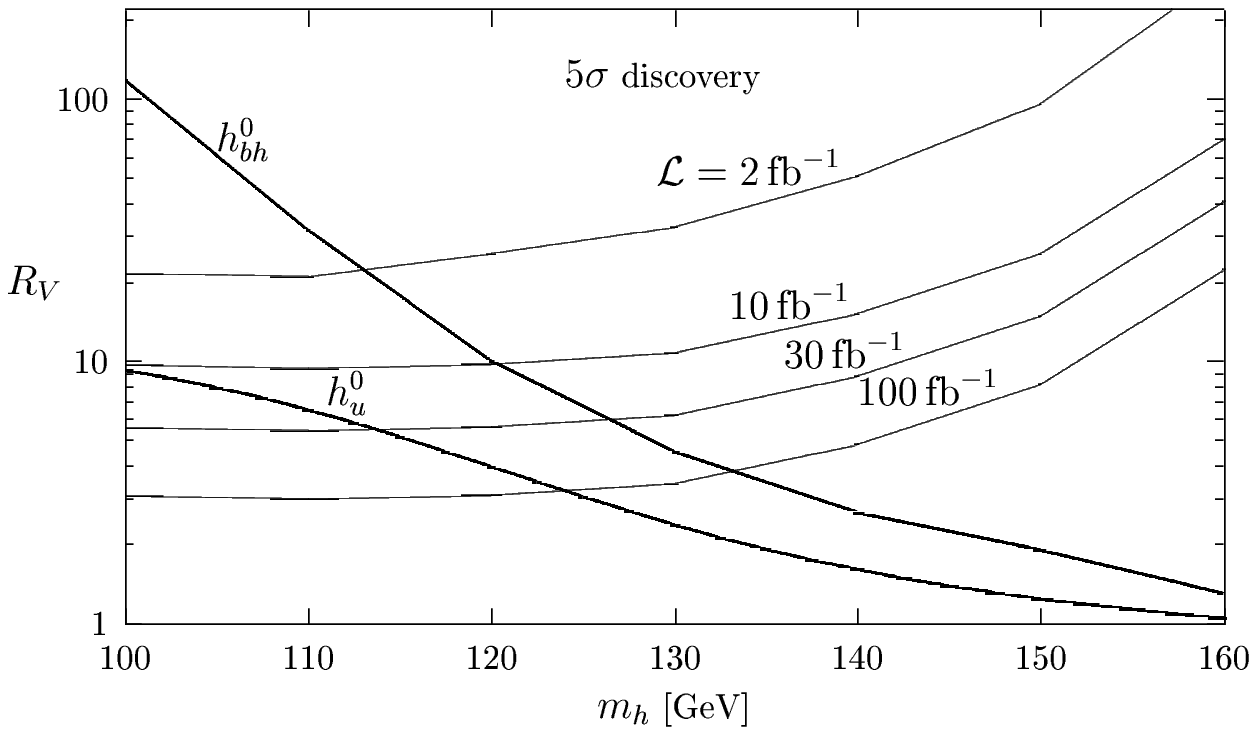}{$5\sigma$ discovery contours of Higgs boson
detection in the $R_{V}$-$m_h$ plane for a given luminosity.  
For each luminosity curve,
the region above the curve can be discovered with Tevatron data.}

%%%%%%%%%%%%%%%%%%%%%%%%%%%%%%%%%%%%%%%%%%%%%%%%%%%%%%%%%%%%%%%
\section*{Discussion and conclusion}

We have analyzed the capability of the Tevatron to find a
Higgs boson decaying into two photons.  We have found that the SM
Higgs boson cannot be probed beyond LEP2 capabilities if the
Tevatron accrues less than $100\xfb^{-1}$.   However, Higgs bosons
in theories beyond the Standard Model may be probed (discovered
or excluded) effectively with significantly less luminosity.
For example, a Higgs boson that couples only to the vector bosons
but has no couplings to the fermions can be probed up to $127\gev$
with less than $\sim 10\xfb^{-1}$ integrated luminosity.  
In the MSSM, when $h \simeq \hu$, so that $h\to b\bar b$ is suppressed
and the $W/Z+h(\to b\bar b)$ signal vanishes,
our analysis shows coverage up to $m_h=114$ GeV with 30 fb$^{-1}$,
and exclusion capability up to $m_h=129\gev$.

In an attempt to be as model independent as possible, we have presented
graphs (Figs.~\ref{gghyy2}-\ref{Vhyy5}) 
of exclusion and detectability as integrated luminosity contours
in the plane of Higgs mass and $R_i$ ($R_{gg}$ and $R_V$), where
$R_i$ parameterizes the enhancement of the $\gamma\gamma$ signal cross-section
over the Standard Model.  Therefore, any theories beyond the SM that have
predictions for production cross-sections and decay widths of Higgs bosons
can be compared with these graphs to attain an estimate of the Tevatron's
capability.  Of course, these figures are not applicable to a Higgs boson that
has an intrinsic width greater than the detector resolution. The theories
discussed above are far from this case.

Finally, we comment on previous studies of $\gamma\gamma$ invariant
mass signals at the Tevatron~\cite{Stange:1994ya}-\cite{Landsberg:2000ht}.
Much
of the earlier work emphasized detectability at lower luminosities of
$h_{\rm bh}^0$ with $m_{h_{\rm bh}^0}<100\gev$,
where the branching fraction to two photons was ${\cal O}(1)$.  
For example, this was the Higgs boson and 
the mass region covered in Ref.~\cite{Stange:1994ya}.
They also performed their simulations at the parton level and applied
much looser cuts than our analysis, and much looser than those
typically used in the experimental analyses
of D\O ~and CDF 
in Run I~\cite{experiment}.

Another important analysis was completed very recently
in Ref.~\cite{Landsberg:2000ht}
with results similar to ours, although
the analysis differs in several ways ({\it e.g.}, no $p_T^{\gamma\gamma}$ 
cut). This study suggests
that a signal of
two photons combined with a single jet or two jets is
an effective method to search for a Higgs boson in the two photon decay mode,
but it did not utilize cuts as stringent as ours.
A careful comparison 
needs to be made among all the observables in the various studies
under precisely the same
assumptions to ascertain which observables are the most effective.  And,
of course, a combination of all useful observables should be employed
to maximize our sensitivity to Higgs bosons.  The suggested two photon
observables outlined in this paper appear to be useful additions to the list
of Higgs boson search observables.

%%%%%%%%%%%%%%%%%%%%%%%%%%%%%%%%%%%%%%%%%%%%%%%%%%%%%%%%%%%%%%%%%%%%%%
\medskip
\noindent
{\it Acknowledgements:} J.W. thanks Lawrence Berkeley National
Laboratory for partial support in the participating guest program.
SM thanks G.L.~Kane for useful conversations.

%%%%%%%%%%%%%%%%%%%%%%%%%%%%%%%%%%%%%%%%%%%%%%%%%%%%%%%%%%%%%%%%%%%%%%%%
\section*{Appendix: estimation of signal and backgrounds}
\subsection*{Signal}
As mentioned in the text, we are mainly concerned with the
processes $gg\to h$ and $q\bar q\to Vh$ (with $V=W$ or $Z$).
The $gg\to h\to\gamma\gamma$ process is calculated based on 
$b$-space resummation (see, {\it e.g.}, Ref.~\cite{Balazs:2000wv}), 
and performed to
NLO accuracy.  The total event rates (without cuts) agree 
with other fixed order calculations~\cite{Spira:1995mt}.
Since the multiple, soft gluon emissions are integrated out,
the effect of isolation cuts must be determined by some
other means.  We use a constant isolation efficiency per
photon $\epsilon_{\rm iso}=0.95$ for these inclusive studies.
Our results can be easily scaled if necessary to account for a
different efficiency.

The $q\bar q\to Vh$ process is calculated using {\sc PYTHIA}~\cite{Sjostrand:1994yb},
but multiplied by a constant $K$--factor based
on the resummed calculation of Ref.~\cite{Mrenna:1998wp}.  
For completeness, the contribution of vector boson fusion
processes were also calculated using {\sc PYTHIA}
without any effective-$W$ approximation and no $K$ factor.  
This process was never
relevant for our analysis, for reasons discussed in the main text.

\subsection*{Background}
The background estimate of the inclusive production of $\gamma\gamma$
pairs -- where kinematic cuts are applied only on the properties of
the individual photons or the diphoton pair -- uses
the next--to--leading order, resummed calculation of Ref.~\cite{Balazs:1998xd}
applied to the 2.0 TeV collider energy.  Since the resummed calculation
integrates out the history of the soft gluon emission, the photon isolation
efficiency must be estimated by another means, such as a showering Monte
Carlo or from $Z$ boson data.  We use a constant isolation efficiency per
photon $\epsilon_{iso}=0.95$ as for the signal.  No backgrounds from fragmentation
photons (e.g. $\pi^0,\eta\to\gamma\gamma$) are included in our numbers.
The results of Ref.~\cite{Balazs:1998xd} show good agreement with Run I 
data, and contain only a small component of fragmentation photons.  For
simplicity, we have ignored it entirely.  
Of course, the actual contribution
from fragmentation photons depends critically on the isolation criteria
and on the minimum $p_T^\gamma$.
Note that the resummed calculation
{\it does} include the final state 
bremsstrahlung processes, {\it e.g.} $qg\to q\gamma\gamma$.  We find that our
calculational method yields good agreement with 
the invariant mass distribution presented in Ref.~\cite{wilson} without
a large fragmentation component.

To estimate the backgrounds to $W$ or $Z$ + $\gamma\gamma$, where the
gauge bosons decay leptonically or hadronically, we need to determine
the properties of the individual quarks and gluons emitted in the standard
$\gamma\gamma$ production processes.  This is not straightforward,
since parton showering is accurate at describing event shapes but
not event rates, whereas the hard NLO corrections to the $\gamma\gamma$
production rate are known to be important.  For moderate values of
$p_T^{\gamma\gamma}$ relative to $M_{\gamma\gamma}$, a fixed order
(in $\alpha_s$) calculation is as accurate in describing the
kinematics of the photon pair as a resummed one (the transition
between the two perturbative schemes is handled naturally in the
resummation formalism, but the gluon emissions are integrated out).  
Therefore, we use the
partonic subprocesses $q\bar q\to\gamma\gamma g$, $qg\to \gamma\gamma q$, 
etc., to set the event rate, plus the parton showering method
to reconstruct the full history of possibly multiple gluon emissions.
For the $gg\to\gamma\gamma$+jets background, we use parton showering with
the $gg\to\gamma\gamma$ process, but using the improvements of 
Ref.~\cite{meshower}
to approximate the NLO corrections (the effect of using the exact pentagon
diagram for the $gg\to\gamma\gamma g$ process is not 
important~\cite{balazs private}).  The hard scale is set to
the photon pair invariant mass. In all cases, we calculate the
isolation efficiency explicitly.

%%%%%%%%%%%%%%%%%%%%%%%%%%%%%%%%%%%%%%%%%%%%%%%%%%%%%%%%%%%%%%%%%%%%%%%%


\begin{thebibliography}{20}

\bibitemx{runII web}
For latest results, see {\tt http://fnth37.fnal.gov/higgs.html}.

%\cite{Djouadi:1998yw}
\bibitemx{Djouadi:1998yw}
A.~Djouadi, J.~Kalinowski and M.~Spira,
``HDECAY: A program for Higgs boson decays in the standard model and its  supersymmetric extension,''
Comput.\ Phys.\ Commun.\  {\bf 108}, 56 (1998)
[hep-ph/9704448].


\bibitemx{deltas}
M.~Carena, J.~R.~Espinosa, M.~Quiros and C.~E.~Wagner,
``Analytical expressions for radiatively corrected Higgs masses and
couplings in the MSSM,''
Phys.\ Lett.\  {\bf B355}, 209 (1995)
[hep-ph/9504316];
M.~Carena, M.~Quiros and C.~E.~Wagner,
``Effective potential methods and the Higgs mass spectrum in the MSSM,''
Nucl.\ Phys.\  {\bf B461}, 407 (1996)
[hep-ph/9508343];
H.~E.~Haber, R.~Hempfling and A.~H.~Hoang,
``Approximating the radiatively corrected Higgs mass in the minimal
supersymmetric model,''
Z.\ Phys.\  {\bf C75}, 539 (1997)
[hep-ph/9609331].

%\cite{Baer:1998cm}
\bibitemx{Baer:1998cm}
H.~Baer and J.D.~Wells,
``Trilepton Higgs signal at hadron colliders,''
Phys.\ Rev.\ {\bf D57}, 4446 (1998)
hep-ph/9710368.
%%CITATION = PHRVA,D57,4446;%%

%\cite{Loinaz:1998ph}
\bibitemx{Loinaz:1998ph}
W.~Loinaz and J.D.~Wells,
``Higgs boson interactions in supersymmetric theories with large $\tan\beta$,''
Phys.\ Lett.\ {\bf B445}, 178 (1998)
hep-ph/9808287.
%%CITATION = PHLTA,B445,178;%%

%\cite{Carena:1998gk}
\bibitemx{Carena:1998gk}
M.~Carena, S.~Mrenna and C.E.~Wagner,
``MSSM Higgs boson phenomenology at the Tevatron collider,''
Phys.\ Rev.\ {\bf D60}, 075010 (1999)
hep-ph/9808312.
%%CITATION = PHRVA,D60,075010;%%


%\cite{Carena:1999bh}
\bibitemx{Carena:1999bh}
M.~Carena, S.~Mrenna and C.E.~Wagner,
``The complementarity of LEP, the Tevatron and the LHC in the search for  a light MSSM Higgs boson,''
hep-ph/9907422.
%%CITATION = HEP-PH 9907422;%%

%\cite{Hall:1994gn}
\bibitemx{Hall:1994gn}
L.J.~Hall, R.~Rattazzi and U.~Sarid,
``The Top quark mass in supersymmetric SO(10) unification,''
Phys.\ Rev.\ {\bf D50}, 7048 (1994)
hep-ph/9306309.
%%CITATION = PHRVA,D50,7048;%%

%\cite{Dine:1997xk}
\bibitemx{Dine:1997xk}
M.~Dine, Y.~Nir and Y.~Shirman,
``Variations on minimal gauge mediated supersymmetry breaking,''
Phys.\ Rev.\ {\bf D55}, 1501 (1997)
hep-ph/9607397.
%%CITATION = PHRVA,D55,1501;%%

\bibitemx{bosonic higgs}
H.~E.~Haber, G.~L.~Kane and T.~Sterling,
``The Fermion Mass Scale And Possible Effects Of Higgs Bosons On Experimental Observables,''
Nucl.\ Phys.\  {\bf B161}, 493 (1979);
J.~F.~Gunion, R.~Vega and J.~Wudka,
``Higgs Triplets In The Standard Model,''
Phys.\ Rev.\  {\bf D42}, 1673 (1990);
B.~Grzadkowski and J.~F.~Gunion,
``Kaluza-Klein excitations and electroweak symmetry breaking,''
hep-ph/9910456.

%\cite{Wells:1997pg}
\bibitemx{Wells:1997pg}
J.D.~Wells,
``The electroweak symmetry breaking Higgs boson in models with top-quark  condensation,''
Phys.\ Rev.\ {\bf D56}, 1504 (1997)
hep-ph/9612292.
%%CITATION = PHRVA,D56,1504;%%

\bibitemx{top condensation}
C.~T.~Hill,
``Topcolor: Top quark condensation in a gauge extension of the standard model,''
Phys.\ Lett.\  {\bf B266}, 419 (1991);
C.~T.~Hill,
``Topcolor assisted technicolor,''
Phys.\ Lett.\  {\bf B345}, 483 (1995)
[hep-ph/9411426].

\bibitemx{Spira:1995rr}
M.~Spira, A.~Djouadi, D.~Graudenz and P.~M.~Zerwas,
``Higgs boson production at the LHC,''
Nucl.\ Phys.\  {\bf B453}, 17 (1995)
[hep-ph/9504378].

%\cite{Kane:1996ek}
\bibitemx{Kane:1996ek}
G.L.~Kane, G.D.~Kribs, S.P.~Martin and J.D.~Wells,
``Two photon decays of the lightest Higgs boson of supersymmetry at the LHC,''
Phys.\ Rev.\ {\bf D53}, 213 (1996)
hep-ph/9508265.
%%CITATION = PHRVA,D53,213;%%

%\cite{Djouadi:1998az}
\bibitemx{Djouadi:1998az}
A.~Djouadi,
``Squark effects on Higgs boson production and decay at the LHC,''
Phys.\ Lett.\  {\bf B435}, 101 (1998)
[hep-ph/9806315].


%\cite{Spira:1997ce}
\bibitemx{Spira:1997ce}
M.~Spira and J.D.~Wells,
``Higgs bosons strongly coupled to the top quark,''
Nucl.\ Phys.\ {\bf B523}, 3 (1998)
hep-ph/9711410.
%%CITATION = NUPHA,B523,3;%%

\bibitem{Hall:1999fe}
L.~Hall and C.~Kolda,
``Electroweak symmetry breaking and large extra dimensions,''
Phys.\ Lett.\  {\bf B459}, 213 (1999)
[hep-ph/9904236].

\bibitem{Abdullin:1998er}
S.~Abdullin, M.~Dubinin, V.~Ilyin, D.~Kovalenko, V.~Savrin and N.~Stepanov,
``Higgs boson discovery potential of LHC in the channel 
$p p\to\gamma\gamma+j$,''
Phys.\ Lett.\  {\bf B431}, 410 (1998)
[hep-ph/9805341].                  

%\cite{Stange:1994ya}
\bibitemx{Stange:1994ya}
A.~Stange, W.~Marciano and S.~Willenbrock,
``Higgs bosons at the Fermilab Tevatron,''
Phys.\ Rev.\  {\bf D49}, 1354 (1994)
[hep-ph/9309294].

%\cite{Akeroyd:1996hg}
\bibitemx{Akeroyd:1996hg}
A.~G.~Akeroyd,
``Fermiophobic Higgs bosons at the Tevatron,''
Phys.\ Lett.\  {\bf B368}, 89 (1996)
[hep-ph/9511347].

%\cite{Abachi:1997dc}
%\bibitemx{Abachi:1997dc}
\bibitem{experiment}
S.~Abachi {\it et al.}  [D0 Collaboration],
``Search for diphoton events with large missing transverse energy in  
p anti-p collisions at $\sqrt{s}= 1.8$-TeV,''
Phys.\ Rev.\ Lett.\  {\bf 78}, 2070 (1997)
[hep-ex/9612011];
%\cite{Abe:1998up}
%\bibitemx{Abe:1998up}
F.~Abe {\it et al.}  [CDF Collaboration],
``Searches for new physics in diphoton events in $p\bar p$ 
collisions at  $\sqrt{s}= 1.8$-TeV,''
Phys.\ Rev.\ Lett.\  {\bf 81}, 1791 (1998)
[hep-ex/9801019];
%\cite{Abe:1999ui}
%\bibitemx{Abe:1999ui}
F.~Abe {\it et al.}  [CDF Collaboration],
``Searches for new physics in diphoton events in 
$p\bar p$ collisions at  $\sqrt{s}= 1.8$-TeV,''
Phys.\ Rev.\  {\bf D59}, 092002 (1999)
[hep-ex/9806034];
%%%% D0 Collaboration
%%\cite{Abbott:1999vv}
%\bibitemx{Abbott:1999vv}
B.~Abbott {\it et al.}  [D0 Collaboration],
``Search for nonstandard Higgs bosons using high mass photon pairs in  
$p\bar p\to \gamma\gamma+2j$ at $\sqrt{s} = 1.8$-TeV,''
Phys.\ Rev.\ Lett.\  {\bf 82}, 2244 (1999)
[hep-ex/9811029].

\bibitem{wilson}
P.J.~Wilson et al. (CDF Collaboration),
``Search for high mass photon pairs in $p\bar p$ 
collisions at $\sqrt{s} = 1.8$ TeV,''
presented at {\it 29th International Conference on High-Energy Physics
(ICHEP 98)}, Vancouver, British Columbia, Canada, July 23-30, 1998
(Fermilab-Conf-98/213-E).

%\cite{Gonzalez-Garcia:1998ik}
\bibitemx{Gonzalez-Garcia:1998ik}
M.~C.~Gonzalez-Garcia, S.~M.~Lietti and S.~F.~Novaes,
``Search for non-standard Higgs boson in diphoton events at 
$p\bar p$  collisions,''
Phys.\ Rev.\  {\bf D57}, 7045 (1998)
[hep-ph/9711446].



\bibitem{Casalbuoni:1999fs}
R.~Casalbuoni, A.~Deandrea, S.~De Curtis, D.~Dominici, 
R.~Gatto and J.~F.~Gunion,
``Detecting and studying the lightest pseudo-Goldstone boson 
at future  $p p$, $e^+ e^-$ and $\mu^+ \mu^-$ colliders,''
Nucl.\ Phys.\  {\bf B555}, 3 (1999)
[hep-ph/9809523].




%\cite{Landsberg:2000ht}
\bibitemx{Landsberg:2000ht}
G.~Landsberg and K.~T.~Matchev,
``Discovering a light Higgs boson with light,''
hep-ex/0001007.
%%CITATION = HEP-EX 0001007;%%
%\href{http://www.slac.stanford.edu/spires/find/hep/www?eprint=HEP-EX/0001007}{SPIRES}

\bibitem{Balazs:2000wv}
C.~Balazs and C.~P.~Yuan,
``Higgs boson production at the LHC with soft gluon effects,''
hep-ph/0001103.

\bibitem{Spira:1995mt}
M.~Spira,
``HIGLU: A Program for the Calculation of the Total Higgs Production Cross Section at Hadron Colliders via Gluon Fusion including QCD
Corrections,''
hep-ph/9510347.

%\cite{Sjostrand:1994yb}
\bibitemx{Sjostrand:1994yb}
T.~Sjostrand,
``High-energy physics event generation with PYTHIA 5.7 and JETSET 7.4,''
Comput.\ Phys.\ Commun.\  {\bf 82}, 74 (1994).
%%CITATION = CPHCB,82,74;%%
%\href{http://www.slac.stanford.edu/spires/find/hep/www?j=CPHCB,82,74}{SPIRES}

%\cite{Mrenna:1998wp}
\bibitemx{Mrenna:1998wp}
S.~Mrenna and C.~P.~Yuan,
``Effects of QCD resummation on $W+ h$ and 
$t\bar b$ production at the  Tevatron,''
Phys.\ Lett.\  {\bf B416}, 200 (1998)
[hep-ph/9703224].

%\cite{Balazs:1998xd}
\bibitemx{Balazs:1998xd}
C.~Balazs, E.L.~Berger, S.~Mrenna and C.P.~Yuan,
``Photon pair production with soft gluon resummation in hadronic  interactions,''
Phys.\ Rev.\ {\bf D57}, 6934 (1998)
hep-ph/9712471.
%%CITATION = PHRVA,D57,6934;%%


\bibitemx{meshower}
S.~Mrenna,
``Higher order corrections to parton showering from resummation  calculations,''
hep-ph/9902471.

\bibitemx{balazs private}
C.~Balazs, private communication.

%>>>>>>>>>>>>>>>>>>>>>>>>>>>>>>>>>>>>>>>>>>>>>>>>>>>>>>>>>>>>>>>>>>


%\cite{Gunion:1997qz}
%\bibitemx{Gunion:1997qz}
%J.F.~Gunion,
%``Detecting and studying Higgs bosons,''
%hep-ph/9705282.
%%CITATION = HEP-PH 9705282;%%


\end{thebibliography}
\end{document}